\documentclass[journal]{IEEEtran}

\usepackage[normalem]{ulem}
\usepackage{comment}

\newcommand{\chk}[1]{{#1}}

\ifCLASSINFOpdf
\else
\fi
 \usepackage[caption=false,font=normalsize,labelfont=sf,textfont=sf]{subfig}
\usepackage{amssymb}
\usepackage{tikz}
\newcommand{\tikzxmark}{%
\tikz[scale=0.23] {
    \draw[line width=0.7,line cap=round] (0,0) to [bend left=6] (1,1);
    \draw[line width=0.7,line cap=round] (0.2,0.95) to [bend right=3] (0.8,0.05);
}}

\usepackage[hidelinks]{hyperref}
\usepackage{graphicx}
\usepackage{listings}
\usepackage{booktabs}
\usepackage{xcolor}
\definecolor{azure}{rgb}{0.0, 0.5, 1.0}
\definecolor{bittersweet}{rgb}{1.0, 0.44, 0.37}
\definecolor{byzantine}{rgb}{0.74, 0.2, 0.64}
\newcommand{\orcid}[1]{\href{https://orcid.org/#1}{\includegraphics[height=10pt]{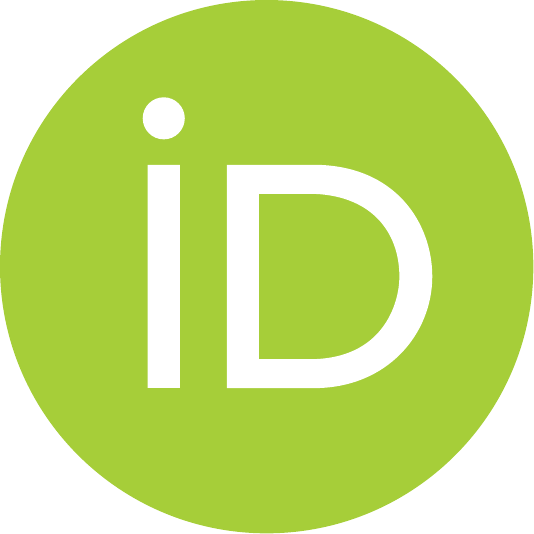}}}

\begin{document}
\title{Locking Down Science Gateways with Landlock and Seccomp}

\author{Steven R Brandt\IEEEauthorrefmark{1}\orcid{0000-0002-7979-2906},
        Max Morris\IEEEauthorrefmark{1}\orcid{0009-0005-8014-1536},
        Patrick Diehl\IEEEauthorrefmark{3}\IEEEauthorrefmark{1}\IEEEauthorrefmark{2}\orcid{0000-0003-3922-8419}, Christopher Bowen\IEEEauthorrefmark{4}, Jacob Tucker\IEEEauthorrefmark{4}, Lauren Bristol\IEEEauthorrefmark{4}, and Golden G. Richard III\IEEEauthorrefmark{1}\IEEEauthorrefmark{4}  \\ 
         \IEEEauthorrefmark{1}LSU Center for Computation \& Technology, Louisiana State University, Baton Rouge, LA, 70803 U.S.A. \\
        \IEEEauthorrefmark{2} Department of Physics and Astronomy, Louisiana State University, Baton Rouge, LA, 70803 U.S.A. \\
        \IEEEauthorrefmark{3} Applied Computer Science (CCS-7), Los Alamos National Laboratory, Los Alamos, NM 87545 U.S.A. \\
        \IEEEauthorrefmark{4} Department of Computer Science, Louisiana State University, Baton Rouge, LA, 70803 U.S.A. 
}

\maketitle

\begin{abstract}
The most recent Linux kernels have a new feature for securing applications: Landlock. Like Seccomp before it, Landlock makes it possible for a running process to selectively relinquish access to certain system resources including network and filesystem access. Rather than being a proper successor to Seccomp, Landlock has different and somewhat complementary capabilities. In this paper, we will evaluate the applicability of using a combination of Landlock and Seccomp (Landlock/Seccomp) in the setting of securing science gateways---codes which need to parse and act upon user-supplied and often unsanitized input. We will explore three case studies in adding Landlock/Seccomp to mature scientific codes: Einstein Toolkit (a code that studies the dynamics of relativistic astrophysics, e.g. neutron star collisions), Octo-Tiger (a code for studying the dynamics of non-relativistic astrophysics, e.g. white dwarfs), and FUKA (an initial data solver for relativistic codes). Finally, we will discuss the implementation of a fully-featured FUKA science gateway designed around Landlock/Seccomp as its primary security mechanism without any user authentication needed.
\end{abstract}

\IEEEpeerreviewmaketitle

\section{Introduction}

Science gateways typically provide a graphical or web interface to scientific code, enabling those users who lack access to or are less savvy with the command line, supercomputers, Slurm~\cite{jette2023architecture}, etc.,
to have ready access to advanced codes. Often, in the interest of democratization, the vetting process for users of the gateway is less rigorous than a typical user account.

To date, as far as we are aware, there are no public, prominently documented cases for which a science gateway or portal was the primary vector of attack. The most closely related incident is the Stakkato attack~\cite{stakkato} in which a Swedish teenager (with help) compromised hundreds of sites worldwide, including supercomputing centers, national labs (DOE, NASA, NCAR), universities, and early grid infrastructure (TeraGrid precursors) by leveraging the implicit trust present in much of the academic environment. Over 1,000 hosts were affected. Methods included credential theft, weak password practices, and propagation across systems, leading to file destruction and root compromises.

Science gateways currently live in a metaphorical ``safe neighborhood'' of relatively insular academic usage, and it is therefore easy to imagine that we are more secure than we really are. It is only a matter of time until a science gateway becomes the vector for a serious attack.

How might such an attack operate? What vulnerabilities do science gateways possess? We would submit that there is one attack surface that is nearly universal among science gateways: user input. Scientific codes are typically written in C, C\texttt{++}, or Fortran by scientists who are domain experts first and programmers second, with a primary focus on results and productivity rather than security. Consequently, these applications represent a potential hazard via buffer overruns, poor input sanitization, etc. Full audits of these codes pose a cost few are willing to undertake. 

An ideal solution for these systems would be to sandbox the code, limiting what it can do even if a hacker were to gain control of the running process. Because these applications typically run in a distributed fashion over MPI, these applications need the ability to make network connections when starting up, but not afterwards. The ideal security configuration would thus be to turn on the sandbox (and take away the ability to make new connections) after calling \texttt{MPI\_Init()} but before processing user-supplied inputs and parameter files. Landlock and Seccomp together (Landlock/Seccomp) can achieve this capability, as well as the ability to impose other typical sandbox restrictions, e.g., what files or directories the process should be able to read and write.

To evaluate the practical value of Landlock/Seccomp in the setting of science gateways, we pose the following questions:
\begin{itemize}
    \item Can Landlock/Seccomp readily address the security needs of the science gateway and make sophisticated scientific applications secure?
    \item Can Landlock/Seccomp accomplish this without overburdening the gateway or application developer with the need to have prior cybersecurity knowledge or make extensive modifications to their code?
\end{itemize}

In this paper, we will apply Landlock/Seccomp to three complex science codes: The Einstein Toolkit, which we will use to simulate a spherically symmetric neutron star; Octo-Tiger, which we will use to simulate a white dwarf; and FUKA, which we will use to generate binary neutron star initial data sets. These codes have little in common except that they are large C\texttt{++} codes in the astrophysics domain. We will show that it is relatively easy to modify these codes to employ Landlock. In our previous work~\cite{oldpaper} on this topic, we focused exclusively on whether it was possible to execute scientific applications with MPI under Landlock. In this work, we augment the use of Landlock with a synergistic use of Seccomp, and strengthen the case for the use of Landlock/Seccomp use as the primary security mechanism for a science gateway by implementing a fully-functional and secure gateway server.

The paper is structured as follows: Section~\ref{sec:related:work} presents the related work. Section~\ref{sec:security:tools} discusses tools and facilities that can be used to address the problem, Section~\ref{sec:scientific:applications} briefly introduces the scientific applications to be studied. Section~\ref{sec:implementation} describes the implementation and testing. Section~\ref{sec:configuration} shows how to lock access to the file system and network. Section~\ref{sec:runtime:measurements} provides run time measurements for our chosen applications with and without \texttt{Landlock}. Finally, Section~\ref{sec:conclusion} concludes the work.

\section{Related work}
\label{sec:related:work}
The security of science gateways is a multi-faceted problem. Much of the existing science gateway security literature has focused primarily on the perimeter: authentication, authorization, and credential delegation. This is unsurprising given the kinds of attacks that have previously taken place, such as the aforementioned Stakkato attack.

To address this, Basney et al.\ developed an OAuth-based delegation framework so that gateway users need not expose long-lived passwords to the gateway itself~\cite{basney2011distributed}. Christie et al.\ extended this line of work in the context of Apache Airavata, unifying multiple credential types across virtual organizations via OAuth2 and OpenID-Connect~\cite{christie2020managing}. Ranawaka et al.\ built Custos~\cite{ranawaka2020custos}, a multi-tenant security middleware layer under the Airavata umbrella that provides federated authentication, identity management, and credential storage through a language-independent API. Trusted~CI, the NSF Cybersecurity Center of Excellence, has published practical guidance on improving science gateway security~\cite{trustedci} and provides cybersecurity consultations to gateway developers through the SGCI Incubator program.

Brandt et al.~\cite{brandtsecuring} addressed the security of the runtime environment itself, developing \texttt{pieshell}, a restricted user-space shell, and a restricted SFTP implementation designed to limit what a gateway’s community account can do on the host system. The thought was to minimize the trust given to the middleware, in this case Tapis/Agave.

The above efforts, while valuable, fail to address the scientific application itself and how it might react to arbitrary input. In some cases, these inputs can be easily restricted or sanitized via the gateway frontend, but many applications require arbitrarily large, complex, and/or binary files as input, e.g., bathymetry files for coastal simulations. Given this huge upper bound of complexity, it is not always viable to parse and sanitize the input. It would be a significant limitation if gateways were not allowed to accept such data. So, what can be done?

Sultan et al.~\cite{sultan2019container} provide a comprehensive survey of container security, organizing Linux kernel confinement primitives---namespaces, cgroups, capabilities, seccomp, and Linux Security Modules---around four threat-model use cases. Their use case~(III), protecting the host from a container, maps closely onto our science gateway problem. Our approach can be seen as a lighter-weight variant that does not require a container runtime.

The Apptainer docs discuss how their security model addresses untrusted users and untrusted containers~\cite{apptainer}. They suggest mounting the container system using the nosuid option, and keeping the \texttt{PR\_NO\_NEW\_PRIVS} flag set. This does not address the possibility of an application making connections to the outside world and attempting exploits on the same machine or local network. While it is, in principle, possible to configure firewalls inside an apptainer image, it is still the case that some applications need to run on MPI connecting different nodes together.

Recently, Multikernel has created Sandlock~\cite{sandlock}, based on similar ideas to those presented in this work, using Linux kernel facilities such as Landlock for isolation rather than a full container~\cite{process}. Their focus is on AI workloads and securing Python applications from launch time; these goals involve different requirements and constraints than securing MPI-enabled science applications, so these works are not interchangeable.

\section{Security Tools and Methodologies}
\label{sec:security:tools}

  Landlock is an unprivileged Linux Security Module, available since kernel
  5.13 (2021), that lets an application impose path-based filesystem
  restrictions on itself and its descendants; TCP bind and connect
  restrictions were added in kernel 6.7 with API 4. Rules are stackable and
  irreversible: a process can only further restrict itself, never relax a
  restriction, which is what makes the mechanism safe to expose to
  unprivileged code. The remainder of this section places Landlock against
  the other Linux confinement primitives a science gateway developer might
  consider.

  Linux offers several primitives for confining a running process, and they
  differ along three axes that matter for a science gateway: who deploys
  them (sysadmin or application), when they take effect (process lifetime
  or a chosen moment), and what they confine (paths, network endpoints,
  syscalls, or resources). Older filesystem primitives are inadequate on
  their own. \texttt{chroot} is not a security boundary: a process
  retaining \texttt{CAP\_SYS\_CHROOT} or an open descriptor to the original
  root can escape via well-known techniques. \texttt{pivot\_root} was
  designed for the kernel's \texttt{init} to switch root filesystems and
  is the basis of container root assembly, but on its own does not
  restrict network, IPC, or syscalls.

  The \texttt{seccomp} facility has existed in Linux since 2.6.12 in a
  strict mode that allows only a handful of system calls; the more
  flexible BPF-based filter mode (\texttt{seccomp-bpf}) used by modern
  sandboxes was added in Linux~3.5 and underlies the sandboxes in Chrome,
  Firefox, and every mainstream container runtime. Its main limitation for
  our purposes is that filter programs cannot safely dereference pointer
  arguments, so a seccomp filter cannot tell whether an \texttt{open()}
  refers to \texttt{/etc/passwd} or to a user's scratch directory. 

When comparing Seccomp and Landlock to each other, we have discovered that they do not provide the same features for sandboxing. Rather, they have some overlap and complementarity which we will discuss in more detail in a later section. For this reason, we are proposing using both features to secure science gateways.

\chk{As an alternative to combining Seccomp and Landlock, one could consider combining Seccomp with a different tool and create the distinct limitations for reading and writing files using SELinux (which requires root access, setting of config files, etc.) or AppArmor which would provide similar functionality to SELinux and would have similar restrictions.}

  \texttt{SELinux} and \texttt{AppArmor} provide system-wide mandatory access control and are fully
  capable of expressing the restrictions a science gateway needs, but every gateway-specific profile
  must be authored and deployed by the cluster's security team, and the application cannot itself
  enable confinement at a chosen point such as the return from \texttt{MPI\_Init()}.

In comparing the use of AppArmor/SELinux to Landlock, it should be noted that these mechanisms are more mature than Landlock and can work with older kernels. It may be that when locking down a science gateway, that maturity will be more important than the convenience and simplicity Landlock affords. There is a tradeoff to be made here, and no one answer is necessarily right for everyone.

Linux namespaces underpin container runtimes such as Apptainer and can provide strong isolation of a sandboxed process from other processes on the system, but there are some issues. Most critically for our use case, namespaces are configured at launch rather than at a point chosen by the application, so they cannot deliver the after-MPI behavior that we require. In addition, the isolation provided by namespaces is not entirely irrevocable; an external process with appropriate privileges in the parent process namespace can modify it. Lastly, there is some complexity in setup; one has to mount shared libraries, \texttt{/proc}, etc., or the application may crash. It should be noted that Landlock is not entirely free from this complexity, requiring that specific and non-obvious directories or file descriptors be allowed past any filesystem filters depending on the needs of the application.

Virtual machines, e.g., VirtualBox, KVM, can be used to isolate untrusted code. VMs create an entire abstract, isolated system which is hosted by another system. An untrusted application running on a VM could potentially gain administrator privileges or run amok in any number of ways without affecting the host system. The challenge with securing science gateways in a VM lies in how to interact with the sandboxed application without creating any external connections that could be exploited by malicious code. For a science code running with MPI, that might mean a cluster of such machines running on an isolated network. While this might be feasible, it is far more complex than using Landlock/Seccomp.

OpenBSD's \texttt{pledge(2)}
\footnote{\url{http://man.openbsd.org/pledge.2}}
and
\texttt{unveil(2)}
\footnote{\url{http://man.openbsd.org/unveil.2}}
are the closest parallel to Landlock, restricting syscall classes and filesystem paths respectively, but OpenBSD is uncommon in HPC environments.

FreeBSD uses Capsicum~\cite{watson2010capsicum}, a more radical approach to sandboxing than Landlock. Once a process calls \texttt{cap\_enter()}, it loses the ability to access file paths entirely. It can only interact with files by means of file descriptors it already has. Calling Capsicum within an MPI process would naturally take away its ability to create new connections, but the loss of access to the file system paths might require bigger changes to the sandboxed application.

On MacOS and iOS, the tool \textit{XNU Sandbox} provides rules to restrict file system and network access. \textit{XNU Sandbox} is derived from \textit{TrustedBSD} security framework~\cite{watsontrustedbsd}.

Against this landscape, Landlock/Seccomp occupies a specific niche: invoked by the unprivileged application itself, applied at a moment of its choosing, with no sysadmin coordination required. In short, because Landlock/Seccomp works on Linux, and because the combination allows us to check all the boxes in Table~\ref{tab:security:overview}, we believe it is the best presently available choice for sandboxing a scientific application in a science gateway.

\begin{table*}[tb]
    \centering
    \caption{Functionality provided by the Linux Kernel or Tools for sandboxing or access control. Adapted from~\cite{salaunlandlock}.
    While all the features below are important for our application, the ability to enable the system at runtime with one simple
    call was central to our requirements.}
    \begin{tabular}{l|cccccc}\toprule
     Approach & Performance & Fine-grained control &  Embedded policy & Unprivileged use & Enable at Runtime & Block UDP\\\midrule
      Virtual machine & \tikzxmark & \tikzxmark & \tikzxmark &\tikzxmark &\tikzxmark & \checkmark \\
      SELinux   &  \checkmark & \checkmark & \tikzxmark & \tikzxmark &\tikzxmark &\checkmark\\
      Namespaces & \checkmark & \tikzxmark & \tikzxmark & \textasciitilde & \tikzxmark  & \textasciitilde \\
      Seccomp & \checkmark & \tikzxmark & \checkmark & \checkmark & \checkmark & \checkmark\\
      Landlock & \checkmark & \checkmark & \checkmark & \checkmark & \checkmark & \tikzxmark \\\bottomrule
    \end{tabular}
    \label{tab:security:overview}
\end{table*}

\subsection{Security analysis}

The Landlock facility is still fairly new to the Linux kernel. We hypothesized that it might be incomplete. Accordingly, we searched for flaws and attempted to bypass Landlock's access-control.

We started with basic attempts to perform filesystem commands like \texttt{ls} in directories
that were not explicitly allowed in the Landlock sandbox. These attempts were
denied, as expected.
We hypothesized that it may be possible to use these unrestricted commands to provide similar functionality to a disallowed
command---for example, using commands \chk{such as \texttt{find},} it is possible to enumerate folder names in disallowed directories by navigating to the directory and iterating through potential folder and file names to determine if they exist. More complex and sophisticated attacks may benefit from the knowledge gained from this approach.

We made several attempts to circumvent Landlock restrictions using \texttt{sudo}.
Our initial attempts to run \texttt{sudo} failed due to a missing shared library (\texttt{libsudo\_util.so.0}), and the container’s ``no new privileges'' flag prevented privilege escalation. Next, before Landlock was initiated, we copied the 
\texttt{sudo} binary into an ``allowed'' directory with the
set UID bit set, so that it would run with root permissions.
Additionally, we copied \texttt{libsudo\_util.so.0} into the same directory as \texttt{sudo}, setting \texttt{LD\_LIBRARY\_PATH=.} to bring the library into scope. Executing \texttt{LD\_LIBRARY\_PATH=. ./sudo ls} at this point \chk{was blocked by Landlock.} We also made an attempt wherein we reset the \texttt{setuid} permissions on
\texttt{sudo} while running inside Landlock (\texttt{chmod 4755 /usr/bin/sudo}) to restore its ability
to run as root, but this was prevented by Landlock’s filesystem
constraints. All of our attempts to use \texttt{sudo} under Landlock failed.  

Attempts to spawn a subshell from within the \texttt{vi} editor with
\texttt{:set shell=/bin/sh} and \texttt{:shell} succeeded in opening an
interactive shell, but commands were still limited according to Landlock
restrictions. This was expected since Landlock protection extends to forked processes.

Landlock's primary weakness is that, despite its ability to lock down TCP and Unix domain sockets, it does not restrict UDP communication in any way. Since Landlock explicitly does not sandbox UDP (which is disclosed in the documentation), it is essential that other means be used to lock down UDP in order to fully secure the environment. While there are many potential ways to do this, we chose Seccomp, which can simply block all UDP access.

In summary, our security audit determined that the present version of Landlock, augmented with Seccomp, is suitable for safely sandboxing a Linux application.

\subsection{Additional security considerations}
  The empirical results in the previous section establish that Landlock prevents many obvious moves
  a compromised gateway job might attempt against the gateway.  A notable exception is restricting
  UDP traffic, since arbitrary UDP traffic is permitted from within the sandbox.  This is important
  because UDP is the natural carrier for several stealthy exfiltration techniques.  For example, DNS
  tunneling encodes data in subdomain labels of queries to attacker-controlled servers and is widely
  documented in real intrusions.  A Landlock-only sandbox does not address any of these.  This
  particular gap can be closed by pairing Landlock with a small Seccomp filter that denies UDP
  socket creation, or with a per-job network namespace or host-level firewall that restricts
  outbound traffic by user.

  Several other classes of attack lie outside what we tested. Any kernel-mode privilege escalation
  potentially defeats Landlock as it defeats every
  Linux Security Module.  The responsibility for mitigating this belongs to the cluster administrators,
  who must keep kernels patched. Landlock does not retroactively close file descriptors that were
  already open when its rules took effect; our placement of \texttt{landlockme()} after
  \texttt{MPI\_Init()} but before any parameter file is read avoids this in practice, but it is a
  property of how the call is positioned rather than an enforcement guarantee.
  
  Because Landlock is
  path-based, bind mounts assembled before \texttt{landlockme()} are potentially problematic.
  Therefore, we use Seccomp to disallow \texttt{mount()} and related calls completely.  
  
  Although Landlock should handle these next two items, newer I/O subsystems such
  as \texttt{io\_uring} and \texttt{bpf} are also not part of our empirical surface and are best
  blocked outright by Seccomp on a science gateway, since few scientific codes need them.

  Resource exhaustion attacks are also outside of our test space.  For example, fork bombs and file descriptor floods remain possible in Landlocked and Seccomped systems, as neither tool restricts \texttt{fork()}, \texttt{clone()}, \texttt{dup()}, et al.  A compromised process could fork exponentially or open file descriptors up to the system limit, starving other jobs or the gateway itself.
  
  While these attacks are not in the principal scope of this work, we note that they can be mitigated by cgroups (specifically the pids controller for fork limits and the files controller in newer kernels) or by ulimit settings (\texttt{RLIMIT\_NPROC}, \texttt{RLIMIT\_NOFILE}).  In multiuser systems, system administrators should enable resource limits of this kind as a general practice.

  \chk{Recent examples of attacks against Linux machines include Copy Fail (CVE-2026-31431) and Dirty Frag (CVE-2026-43284 / CVE-2026-43500). These attacks write directly into the page cache of any readable file such as \texttt{/etc/passwd} or \texttt{/etc/sudoers}. Landlock can do nothing to prevent these attacks, so it is still necessary for the cluster administrators to apply kernel patches as stated above. However, it is notable that even for these exploits, the Landlock/Seccomp security mechanisms trigger an irrevocable set of restrictions for the process and its descendants, so they should continue to enforce restrictions on opening sockets and/or files, meaning that attackers potentially need to work harder to benefit from an exploit.}

  Finally, many real-world sandbox failures are policy mistakes rather than kernel attacks: a
  writable directory that overlaps with \texttt{LD\_LIBRARY\_PATH}, an allow-listed binary that is
  itself a shell, or an allow-listed TCP endpoint that can be pivoted through.  These must fall onto
  the cluster operator, similar to active kernel patching.

  In short, Landlock is best deployed as one layer of a defense-in-depth stack with a Seccomp filter to
  cover UDP and dangerous syscalls and active kernel patching to avoid escalation attacks.

\section{Scientific applications}
\label{sec:scientific:applications}




The scientific codes we have chosen follow a standard workflow used by many others: initialize MPI, then read user input in the form of parameter and/or data files, and finally produce a result. Generally, any exploits inherent to the application will use the user input as an attack vector, either by exploiting a low-level vulnerability such as a buffer overflow, or a higher-level vulnerability which tricks the application into accessing sensitive resources or executing malicious instructions. To secure such an application, one inserts calls to \texttt{landlockme()} and \texttt{seccompme()}, or the equivalent, before any user input is processed. For this strategy to be effective, the gateway should not give the user any control over command line arguments to the application, as this would supply user input to the application before Landlock can be activated. The insertion of these calls can be performed either by directly editing the source code, or by using PMPI to override \texttt{MPI\_Init()}.


\subsection{The Einstein Toolkit}
The Einstein Toolkit (ET)~\cite{etk} is a hybrid code constructed from C, C\texttt{++}, and Fortran. Its core infrastructure was first created in 1997 and it has been under continuous development since. While the core infrastructure, Cactus, is generic and could be used for any Cauchy problem, the family of science-specific modules in the ET centers on fully relativistic astrophysical simulations, e.g., black holes, neutron stars, supernovae, and cosmology. Cactus provides adaptive mesh refinement (AMR) with subcycling in time. 

We note here that Landlock locks a single OS thread, while Seccomp locks a single process. Accordingly, \texttt{landlockme()} must be called multiple times within a process while \texttt{seccompme()} need only be called once. See Listing~\ref{lst:cactuslock}.

\begin{lstlisting}[float=*,caption={Calling Landlock/Seccomp from Cactus},label={lst:cactuslock}]
#pragma omp parallel
{
  if(landlockme() != 0) abort();
}
if(seccompme() != 0) abort();
\end{lstlisting}

The test problem we are using in this paper is a TOV star~\cite{oppenheimer1939massive}. This is a simple, spherically symmetric neutron star which we model on a full 3D Cartesian grid. While this is more computational infrastructure than is needed for such a simple simulation, it is a common test problem that is run to verify code correctness and to teach students about neutron stars and the ET code. Importantly, the TOV star will exercise all important components of the solvers required for more sophisticated problems.

\subsection{Octo-Tiger}
Octo-Tiger is an astrophysical code designed for simulating the evolution of non-relativistic star systems using adaptive octrees~\cite{marcello2021octo}. Octo-Tiger simulates gravity using a fast-multipole method (FMM), and the hydro equations are solved using a finite volume method with a fully adaptive mesh refinement (AMR) without subcycling in time (subcycling is avoided because of the need to solve elliptic equations). Octo-Tiger is implemented in C\texttt{++} using HPX~\cite{kaiser2020hpx}, a C\texttt{++} standard library for parallelism and concurrency.

Because Octo-Tiger uses HPX for parallelism instead of OpenMP, the code to run on every process and every thread is somewhat more complex, although adding it was just as straightforward. See Listing~\ref{lst:hpxlock}. One reason for the added complexity is that \texttt{hpx\_main()} runs only on a single process and must call all the others to have them run \texttt{landlockme()}.

\begin{lstlisting}[float=*,caption={Calling Landlock/Seccomp from HPX},label={lst:hpxlock}]
void locker() {
    std::size_t n = hpx::get_os_thread_count();
    std::vector<hpx::future<int>> futs;
    futs.reserve(n);

    // Run on every os thread in the current process
    for (std::size_t i = 0; i < n; ++i) {
        hpx::execution::parallel_executor exec{
            hpx::threads::thread_priority::high,
                hpx::threads::thread_stacksize::default_,
                hpx::threads::thread_schedule_hint{static_cast<std::int16_t>(i)}
        };
        futs.push_back(hpx::async(exec, []() { return landlockme(); }));
    }
    bool locked = true;
    // wait for all threads to complete, checking that they all succeeded
    for(auto&& f : futs) {
        if(f.get() != 0) locked = false;
    }
    if(seccompme() != 0)
        locked = false;
    if(locked)
        std::cout << "Secured: " << getpid() << std::endl;
}

HPX_PLAIN_ACTION(locker, locker_action);

int hpx_main(int argc, char* argv[])
{
    std::vector<hpx::future<void>> futures;

    // send a locker task to every process
    for (auto const& loc : hpx::find_all_localities())
      futures.push_back(hpx::async<locker_action>(loc));

    // wait for all locker tasks to complete
    hpx::wait_all(futures)

    // The application code follows
\end{lstlisting}

\subsection{FUKA}
The Frankfurt University/Kadath (FUKA) initial data solver is a collection of initial data solvers based on the eXtended Conformal Thin-Sandwich (XCTS) formulations of Einstein's equations~\cite{papenfort2021new}. From that collection, we begin by providing access to the binary neutron solvers. The Einstein Toolkit is capable of consuming initial data generated by FUKA and evolving it forward in time.

FUKA is a spectral code, a fundamentally different kind of solver than either the Einstein Toolkit or Octo-Tiger. While it still uses MPI for parallelism, it employs different numerical algorithms. FUKA is a memory-hungry application, requiring up to 300GB of main memory to solve a single initial data problem.


\section{Implementation and Testing}
\label{sec:implementation}

We began by testing very basic MPI codes that exchange simple messages of random data using \texttt{MPI\_Send} and \texttt{MPI\_Recv} in order to verify whether our method works.

We were able to show that if \texttt{landlockme()} and \texttt{seccompme()} were called before \texttt{MPI\_Init}, then the application did not run. If we called \texttt{landlockme()} after \texttt{MPI\_Init}, then the application ran without difficulty using MPICH when using the default Fedora 40 implementation. When we attempted the same test using OpenMPI on that platform, \texttt{Landlock} blocked an attempt to use shared memory. A similar test on Ubuntu 24.04 with OpenMPI, however, was successful.

In principle, one can ask OpenMPI not to use shared memory (by setting \texttt{export OMPI\_MCA\_btl={\textasciicircum}sm}). For simplicity, we tested our scientific codes on Fedora using MPICH and on Ubuntu using OpenMPI.

\subsection{The Einstein Toolkit}

The modification to the Einstein Toolkit was straightforward. We were able to identify the function call \texttt{CCTKi\_InitialiseCactus} and insert the call to \texttt{landlockme()} after the call to a method named \texttt{CCTKi\_InitialiseDataStructures}. With the addition of this single line of code, our TOV star example was able to run and generate data files. 

As a double check that \texttt{Landlock} was indeed active, we ran the tests again, running it in directories it was not supposed to be able to access. As expected, \texttt{Landlock} prevented it from reading or writing files.

\subsection{Octo-Tiger}
For the Octo-Tiger version without networking, we called \texttt{landlockme()} and \texttt{seccompme()} immediately in the main method. As expected, \texttt{Landlock} prevented the program from reading or writing files. To enable networking, we had to ensure that each MPI rank called \texttt{landlockme()} and \texttt{seccompme()} in its initialization routine. Adding \texttt{Landlock} to Octo-Tiger was straightforward and had no major issues. While we did not need to make any significant code changes to Octo-Tiger to support Landlock/Seccomp, we did need to make code changes to compile Octo-Tiger with the newer compiler (GCC 14) provided by Fedora 40.
HPX 1.10 is required to have GCC 14 compiler related fixes we patched.

\subsection{The FUKA Gateway}

\begin{figure}
    \centering
    \includegraphics[width=0.99\linewidth]{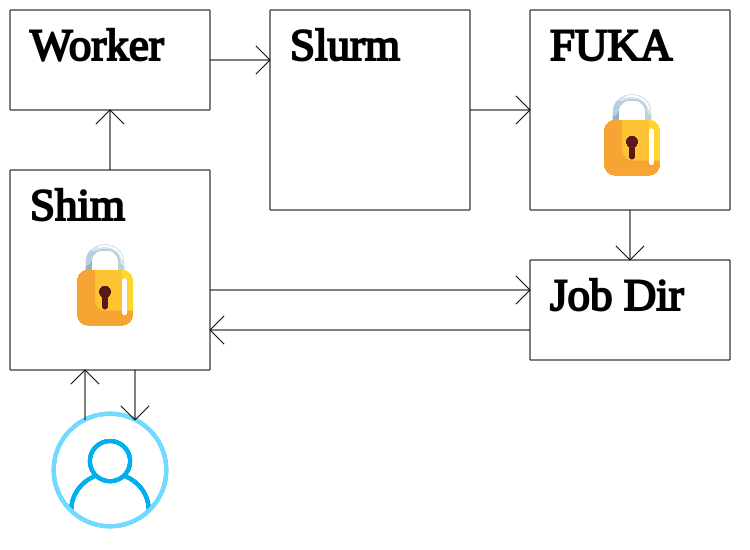}
    \caption{The user interacts with the Shim over HTTP. The Shim parses the user's files, turns the data into searchable key-value pairs and sends it to the worker. The worker then asks Slurm to start the job. Slurm, in turn, starts the FUKA job, which writes to its job directory. Both the Shim (the only attack surface visible to the outside), and the FUKA job (the process which must consume the user data) run under Landlock.}
    \label{fig:fuka-arch}
\end{figure}


In the two previous case studies, we focused only on enabling Landlock/Seccomp in a local execution of a scientific application. For the final case study presented in this paper, we opted to implement FukaGateway~\cite{FukaGateway}, a fully-functional science gateway server around the FUKA initial data solver. FukaGateway establishes a solid, practical proof of concept that Landlock/Seccomp can be used as a security layer for a science gateway in the real world. We chose FUKA in particular for this purpose as its heavy memory requirements (300~GB for a single solution) render it impossible to run on consumer hardware and thus inaccessible to many people in absence of a science gateway. 

FukaGateway exposes REST endpoints over HTTP that respond with structured JSON; a purpose-built web page can be made to facilitate access to the gateway, but simple \texttt{curl} commands work just as well. Users can submit jobs, monitor the progress of their jobs, and download their results. The available endpoints are described in Table~\ref{tab:fuka-endpoints}. 
FukaGateway manages its job queue with Slurm; this is a plausible real-world arrangement. Unlike for the other two applications, we chose to host FukaGateway on the Ubuntu 25.04 operating system, because we were more familiar with how to configure Slurm on that platform.

\begin{table*}
    \centering
    \caption{REST endpoints of the Fuka Gateway server.}
    \begin{tabular}{|c|c|c|}\hline
        \textbf{Operation} & \textbf{Path} & \textbf{Returns}  \\\hline
        Submit Job & \texttt{POST fuka.cct.lsu.edu/jobs} & The job ID  \\\hline
        List Jobs & \texttt{GET fuka.cct.lsu.edu/jobs} & List of all job IDs  \\\hline
        Query Job Status & \texttt{GET fuka.cct.lsu.edu/jobs/:jobId/status} & The status of the job (e.g., running, finished, stopped)   \\\hline
        Query Job Parameters & \texttt{GET fuka.cct.lsu.edu/jobs/:jobId/info}  & The original infofile submitted for a job.  \\\hline
        Download Job Result & \texttt{GET fuka.cct.lsu.edu/jobs/:jobId/result} & The tarball containing the output from a finished job. \\\hline
    \end{tabular}
    
    \label{tab:fuka-endpoints}
\end{table*}

\begin{lstlisting}[float=*,caption={Submitting a job},label={lst:submit}]
    curl -X POST fuka.cct.lsu.edu/jobs --data-binary @initial_bns.info
\end{lstlisting}

Submitting a job to the FUKA Binary Neutron Star (BNS) solver requires a \texttt{initial\_bns.info} infofile. See Listing~\ref{lst:submit}. This is a standard Boost property map describing the positions, separation, spins, masses of the neutron stars, equations of state, etc. When a job is submitted, FukaGateway first parses the infofile to extract the parameters therein. These parameters are matched against previously submitted jobs to determine whether an identical job already exists. If so, the ID of the prior job is returned, and FUKA is never invoked. Otherwise, a job is queued on the nodes using the \texttt{slurm} job manager, and a new ID is returned. The job ID can be passed to various endpoints which operate on individual jobs, e.g., polling the job status or downloading the results from a finished job. See Listings~\ref{lst:poll} and~\ref{lst:download} respectively.

It should be noted that parsing the infofile in the FukaGateway is not necessitated by the use of Landlock/Seccomp and does not serve any security function; it is done only for the purpose of retrieving cached results from previous jobs run with the same parameters. In situations where parsing the user input outside of the science application itself is infeasible, there are no negative implications for security or the use of Landlock/Seccomp. 

\begin{lstlisting}[float=*,caption={Polling a job},label={lst:poll}]
    curl fuka.cct.lsu.edu/jobs/c88afe38-8dc1-40e4-848c-73578417b142/status
\end{lstlisting}

\begin{lstlisting}[float=*,caption={Downloading the results of a job},label={lst:download}]
    curl -o data.tgz fuka.cct.lsu.edu/jobs/56036ff8-e57e-4a04-b97e-bc3b8ca3946f/result
\end{lstlisting}

The FukaGateway server is written in Rust to benefit from its strong memory safety guarantees, minimizing the potential attack surface for common RCE or privilege escalation methods, including buffer overflow and use-after-free exploits. The server is split into two parts: the ``worker'' and the ``shim''. The shim listens on port 80 and acts as a layer between the outside world and the worker. It is responsible for parsing the infofiles and forwarding requests to the worker. The worker, in contrast, only accepts connections from localhost. It is responsible for communicating with Slurm, managing jobs, and maintaining a simple database to enable parameter-based search over existing jobs. The shim runs under Landlock/Seccomp, with TCP restricted to bind on port 80 (HTTP) and connect on port 48879 (the worker), and filesystem access restricted to read-only access to a few critical networking- and threading-related files listed in Table~\ref{tab:shim-files}. Thus, even in the highly unlikely event that a malicious user is able to craft a file which does bad things to a Rust program, they are still unable to cause significant mischief.

Theoretically, the system described above is still vulnerable to denial-of-service attacks. However, the result of such an attack is that the FUKA server becomes unavailable for a time, and nothing worse. DDoS-style attacks, which inundate the server with excessive network traffic, could be prevented with an established DDoS mitigation service. Other denial-of-service attacks, for instance submitting a job with superficially valid parameters crafted to make FUKA misbehave (crash, run slowly, consume excessive resources) are outside the scope of Landlock/Seccomp and require interventions elsewhere---as discussed in Section \ref{sec:security:tools}, Landlock/Seccomp ought to be just one layer in a larger security stack.

Applying Landlock/Seccomp to the worker server proved to be impossible within our time constraints due to the menagerie of permissions required by Slurm. Since invocations of Slurm are, unavoidably, child processes of the worker, they would by design inherit the same Landlock/Seccomp restrictions placed upon the worker. With further work, it may be the case that a working combination of Landlock rules for Slurm will be found. We maintain that the separation between the worker and the shim, the Landlock/Seccomp restrictions placed onto the shim, the isolation of the worker, and the architecture of both programs render the FukaGateway sufficiently secure.

\begin{table}
    \centering
    \begin{tabular}{c}
        \texttt{/proc/self/cgroup}\\
        \texttt{/sys/fs/cgroup}\\
        \texttt{/sys/devices/system/cpu/online}\\
        \texttt{/proc/stat}\\
        \texttt{/etc/hosts}\\
        \texttt{/etc/resolv.conf}\\
        \texttt{/etc/host.conf}\\
        \texttt{/etc/nsswitch.conf}\\
    \end{tabular}
    \caption{Files to which read-only access is required by the shim server.}
    \label{tab:shim-files}
\end{table}


Our deployment of FukaGateway runs on a virtual machine inside a Dell EMC R740 Poweredge system. The CPUs are Intel Xeon Silver 4116 2.1G, 12C/24T. Our virtual machine was configured with 60GB of main memory, 32 CPU cores, and 4TB of disk.

\subsection{Generic Science Codes}

We note that, with Landlock/Seccomp, it is theoretically possible to create a service which runs arbitrary MPI codes on behalf of unknown users. One way to accomplish this would be to accept a binary in the form of a shared library (\texttt{.so} file) with some kind of standard method, e.g., \texttt{gateway\_main()}. The service would call \texttt{MPI\_Init}, then \texttt{landlockme()} and \texttt{seccompme()}, then it would use \texttt{dlopen()} and \texttt{dlsym()} to access and run the user method. By using \texttt{dlopen()} instead of linking the shared object file, we circumvent the potential problem of constructors being called prior to \texttt{landlockme()} in \texttt{C++}'s initialization sequence.

\section{Configuration}
\label{sec:configuration}
Listing~\ref{listing:bashrc} shows some of the configuration options we used in this study. The first option is \lstinline[language=bash,basicstyle=\ttfamily]{LL_FS_RO}, which takes a list of paths separated by colons. These are the files and directories the application is allowed to read. We gave access to the system-wide installed libraries and executables. The second option \lstinline[language=bash,basicstyle=\ttfamily]{LL_FS_RW} provides a list of files and directories the application is allowed to write to. The third option \lstinline[language=bash,basicstyle=\ttfamily]{LL_TCP_BIND} restricts the port binding and the fourth option \lstinline[language=bash,basicstyle=\ttfamily]{LL_TCP_CONNECT} restricts the ports for connections. We refer to the Linux kernel documentation~\cite{linux-kernel} for more options.

One possible danger is that a malicious user could destroy previously computed data in the validly assigned write directories. This can be mitigated by creating a new directory for each job and granting the associated instance of the application write access only to that directory. 

\begin{lstlisting}[language=bash,keywordstyle=\color{azure},basicstyle=\ttfamily,caption={Example variables to Landlock applications.},float=tp,label=listing:bashrc,escapechar=|,stringstyle=\ttfamily\color{byzantine}]
export LL_FS_RO="/bin:/lib/:|\textcolor{byzantine}{\$}|HOME/"
export LL_FS_RW="|\textcolor{byzantine}{\$}|HOME/"
export LL_TCP_BIND=""
export LL_TCP_CONNECT=""
\end{lstlisting}

\section{Runtime measurements}
\label{sec:runtime:measurements}
Landlock is not supposed to introduce any overhead to a running process~\cite{singh2026experimental}. To verify this claim, we compiled Octo-Tiger using Spack~\cite{Gamblin_The_Spack_Package_2015}, then configured Landlock to restrict Octo-Tiger's filesystem access to the user's home directory (for job i/o) and the Spack installation directory. We executed the rotating star problem from Octo-Tiger's test suite with networking for ten steps with four adaptive mesh refinements. We did not observe any Landlock-induced overhead; runs took around 92 seconds, whether Landlock was enabled or not.


\section{Conclusion}
\label{sec:conclusion}
Landlock and Seccomp are promising technologies when it comes to securing scientific applications for use in science gateways, especially when used in combination as Landlock/Seccomp. Because many science gateways make large, mature, complex C/C\texttt{++} and Fortran codes available as open source on the web, they pose a potentially severe security hazard. A poorly sanitized input could be found and used to generate a buffer overrun attack gaining root on the target system.

These scientific codes are usually based on MPI and follow a pattern of starting up, reading initial data files, computing, then generating results. We have demonstrated that modifying such codes to invoke Landlock/Seccomp after MPI startup but before the reading of parameter files effectively secures the code against exploits that rely on user input as an attack vector. We note that Landlock/Seccomp provides no protection against denial of service attacks, e.g., excessive usage of disk space, inodes, file descriptors, etc. While these types of threats can still cause significant problems, they are of a different class and can be mitigated by other established means.

Because Landlock does not support the restriction of UDP activity and may be vulnerable to attacks using \texttt{mount()}, we have supplemented our use of Landlock with Seccomp.
Our findings show that when combined with Seccomp, the version of Landlock in Fedora 40 (API 4) is a good candidate for securing these sorts of applications.

We have also demonstrated that, even with large, complex codes such as Octo-Tiger, the Einstein Toolkit, and FUKA, sandboxing an application with Landlock is a relatively straightforward task. Not only was this done with relative ease, but it introduced no performance penalties or observable runtime overheads. We have further provided a functional proof-of-concept implementation for a Landlock/Seccomp-first science gateway written in Rust. With additional development, we expect this simple server, or servers written with this same structure, to become a useful tool among relativistic scientists.


%

\appendices


\section*{Disclaimer}
This article is an extended version of a paper previously published in the proceedings of Science Gateways 2024 (SG24), titled “Locking Down Science Gateways.” The current manuscript includes significant additions and revisions, such as expanded methodology, the inclusion of Seccomp, additional experiments, the FukaGateway implementation, and a more comprehensive analysis of the results.

\section*{Declarations}
\subsection{Conflict of Interest Statement}
On behalf of all authors, the corresponding author states that there is no conflict of interest. 

\subsection{Ethical Approval}
This declaration is not applicable.

\subsection{Funding}
We acknowledge the support of NSF grant OAC 2004157 to support work on the Einstein Toolkit.

\subsection{Author Contributions}
\begin{itemize}
    \item \textbf{Steven Brandt:} Steven developed the high-level concept for the curl-based gateway and composed much of the text. Steven also configured the Ubuntu server and slurm, and was responsible for modifying and compiling both the Einstein Toolkit and FUKA.
    
    \item \textbf{Max Morris:} Max was responsible for designing, implementing, and testing the FukaGateway server. Max contributed sections of the text relating to FukaGateway, as well as smaller text contributions and revisions throughout the rest of the document.

    \item \textbf{Patrick Diehl:} Patrick was responsible for modifying and compiling Octo-Tiger and running the performance tests with and without. He also contributed to the text of the Related Work section.

    \item \textbf{Christopher Bowen, Jacob Tucker, Lauren Bristol, and Golden G. Richard III:} These authors researched potential methods of attacking our security model and Landlock. They jointly contributed to the text of the {Security Analysis} subsection. Golden also contributed to the {Security Tools and Methodologies} section.
\end{itemize}

\subsection{Availability of data and materials}
\begin{itemize}
    \item Octo-Tiger is available on GitHub~\cite{octotiger} and can be compiled with Spack~\cite{octo-spack}. The scripts and input data to reproduce the runs in Section~\ref{sec:runtime:measurements} are available on Zenodo~\cite{diehl_2024_11355929}.
    \item The Einstein Toolkit is free under the GPLv3 license and is available for public download using instructions found at the Einstein Toolkit website~\cite{einstein}.
    \item FukaGateway is available on GitHub~\cite{FukaGateway} and can be compiled with Rust version \texttt{1.89.0}.
\end{itemize}

\section*{Acknowledgment}

The authors would like to thank the IT support staff of the Center for Computation and Technology who setup our test machines for us. This work was supported by the U.S. Department of Energy through the Los Alamos National Laboratory. Los Alamos National Laboratory is operated by Triad National Security, LLC, for the National Nuclear Security Administration of U.S. Department of Energy (Contract No. 89233218CNA000001). LA-UR-25-29474

\ifCLASSOPTIONcaptionsoff
  \newpage
\fi



\bibliographystyle{IEEEtran}
\bibliography{sn-bibliography}
%

%








\end{document}